# Covariance among independent variables determines the overfitting and underfitting problems in variation partitioning methods: with a special focus on the mixed co-variation

Youhua Chen

Department of Renewable Resources, University of Alberta, Edmonton, T6G 2H1, Canada

Email: haydi@126.com

ABSTRACT

The effectiveness and validity of applying variation partitioning methods in community ecology has been questioned. Here, using mathematical deduction and numerical simulation, we made an attempt to uncover the underlying mechanisms determining the effectiveness of variation partitioning techniques. The covariance among independent variables determines the under-fitting and over-fitting problem with the variation partitioning process. Ideally, it is assumed that the covariance among independent variables will be zero (no correlation at all), however, typically there will be some colinearities. Therefore, we analyzed the role of slight covariance on influencing species variation partitioning. We concluded that when the covariance between spatial and environmental predictors is positive, all the three components-pure environmental, spatial variations and mixed covariation were over-fitted, with the sign of the true covariation being negative. In contrast, when the covariance is negative, all the three components were under-fitted with the sign of true covariation being positive. Other factors, including extra noise levels, the strengths of variable coefficients and the patterns of landscape gradients, could reduce the fitting problems caused by the covariance of variables. The conventional calculation of mixed covariation is incorrect and misleading, as the true and estimated covariations are always sign-opposite. In conclusion, I challenge the conventional three-step procedure of variation partitioning, suggesting that a full regression model with all variables together is robust enough to correctly partition variations.

INTRODUCTION

In ecological communities, one principal process regulates frequently and determines community structure. It is important to consider which kinds of ecological processes are dominant, while others auxiliary. Thus, the variance in response variables can be separated into several parts, and by employing statistical methods, we can identify the contribution and relative importance of different ecological mechanisms.

The characteristic of species composition influenced by environmental variables is a major topic in current ecological research. Redundancy analysis (suited for linear relationships between species composition and environmental variables) and Canonical correspondence analysis (handling nonlinear species-environment relationship) are the widely used methods to investigate the relationship of environmental variables and species diversity information. Variation partitioning can be used to test and determine the possibilities of individual predictors in influencing species distribution and abundance (Peres-Neto et al., 2006). Of particular importance in ecology is the separation of spatial (interpreted as dispersal limitation) and environmental (interpreted as niche limitation) effects on species compositions. Variation partitioning could help resolve this issue largely.

Partitioned variance can be divided into four parts: pure environmental variation, pure space variation, mixed environmental and space variation, and unexplained variation (Borcard et al., 1992). Fig. 1 decipts the four parts of variance. When setting spatial descriptors as covariables and removing their effects, we can know the proportion of pure environmental variation. In contrast, when setting environmental variables as covariables, we can deduce the proportion of pure space variation. The mixed environmental and spatial variation can be derived from the subtraction of pure environmental variation and pure spatial variation from the total explained variance. This is the typical three-step variation partitioning procedure introduced in previous literature (Borcards et al., 1992; Legendre and Legendre, 1997).

However, the variation partitioning may suffer severe fitting problems. A recent study (Gilbert and Bennett, 2010) used simulations to compare the power and accuracy of a variety of variation partitioning methods. They found that all kinds of available multivariate tools have greatly underestimated each of the three parts of variances. Further, there are many debates on the effectiveness of using variation partitioning to reconcile the contribution of niche and neutrality mechanisms in structuring ecological communities (e.g., Smith and Lundholm, 2010; Tuomisto et al., 2012).

Therefore, some critical questions have become natural for us to address: can we accurately estimate the variations caused by pure spatial factors, pure environmental factors and the combination of both? Under what kinds of conditions, we can have the correct estimation and partition of the variation? How can we avoid under-fitting or over-fitting problems?

MATERIALS AND METHODS

Fitting problems in variation partitioning

For a full simle two-variable model, we can write down the equation as,

$$Y = aS + bE + \delta$$

Here $\delta \sim N(0,1)$. $a$ and $b$ are the real coefficients for spatial and environemntal variables respectively to generate the response variable Y. Of course, the model can be extended to matrix form considering multiple variables without losing generality.

The total explained variation therefore, should be,

$$[S+E]_{est} = Var(\hat{Y}_{S \cup E}) \approx [S+E]_{real} = Var(Y_{S \cup E}) = a^2 Var(S) + b^2 Var(E) + 2ab Cov(S,E)$$

Hence, the estimated total explained variation is identical to real total predictable variation due to the constraint that the response variable Y is completely bounded by S and E.

Based on the three-step procedure of variation partitioning, the estimated coefficients for spatial and environmental variables are as follows (see appendix for details),

$$\begin{cases} \hat{a} = a + b \dfrac{Cov(S,E)}{Var(S)} \\ \hat{b} = b + a \dfrac{Cov(S,E)}{Var(E)} \end{cases} \quad (1)$$

Therefore,

When $Cov(S,E) > 0$, we will have $\begin{cases} \hat{a} > a \\ \hat{b} > b \end{cases}$; when $Cov(S,E) < 0$, we have $\begin{cases} \hat{a} < a \\ \hat{b} < b \end{cases}$.

We can deduce the form for mixed variations under true and estimated scenarios as follows, For estimated models, we will have,

$$CoVar(\hat{Y}_{S \cap E}) = [SE]_{est} = \hat{a}^2 Var(S) + \hat{b}^2 Var(E) - (a^2 Var(S) + b^2 Var(E) + 2ab Cov(S,E))$$

However, in real scenarios, we have,

$$CoVar(Y_{S \cap E}) = [SE]_{real}$$
$$= a^2 Var(S) + b^2 Var(E) - (a^2 Var(S) + b^2 Var(E) + 2ab Cov(S,E))$$
$$= -2ab Cov(S,E)$$

There, we found that, it is very hard to accurately estimate the real mixed-variance (only in special cases, e.g., $Cov(S,E) = 0$)

Therefore, the mixed variance difference between real model and estimated model should be,

$$CoVar(\hat{Y}_{S \cap E}) - CoVar(Y_{S \cap E})$$
$$= (\hat{a}^2 Var(S) + \hat{b}^2 Var(E)) - (a^2 Var(S) + b^2 Var(E))$$
$$= (\hat{a}^2 - a^2) Var(S) + (\hat{b}^2 - b^2) Var(E)$$

When $a$ and $b$ were over-estimated (or negative) in the condition of $Cov(S,E) > 0$, the mixed covariation derived from S and E was over-estimated (or negative) as well ($CoVar(\hat{Y}_{S \cap E}) > CoVar(Y_{S \cap E})$).

**Numerical simulations**

Centralization of the sample is a necessary step for performing multivariate statistics, which is important to remove random effects and standardize data as the sample derived from the normal distribution with zero means. Therefore, without mention, all the analyses below are performed on centralized data.

Because most of variation partitioning methods are similar, we only considered the simplest method- redundancy analysis (in our model of course, there is only one response variable, thus the method was reduced as a common linear regression). For each model, both a simple linear regression method and a general additive model were used. Decomposition of variance for each part of the data is shown in Appendix 1.

For generating spatial and environmental structure of the landscape, we consider two simple forms, power form and sin form, which corresponds the spatial and environmental gradients respectively (hereafters, $\delta$-related parameters denote independent White Gaussian noises).

$$S(x) = x^\alpha + \delta_1$$

$$E(x) = \sin(x/\beta) + \delta_2$$

Here x denotes the locations. $S(x)$ and $E(x)$ denotes the spatial and environmental attributes across the landscape.

The species abundance ($Y$) across the landscape, can be assumed as the form of linear combination of spatial and environmental variables, thus, we can write as,

$$Y(x) = aS(x) + bE(x) + \delta$$

Because our simulation actually concerned only the variation splitting of species abundance contributed by spatial and environmental constraints, we thus omit the landscape parameter x from further equations.

$$Y = aS + bE + \delta$$

Here, Y, S, and E were all centralized before performing multiple regression analysis.
The adjusted-$R^2$ metric is employed to assess the explained variations as follows,

$$R^2_{adj} = 1 - (n-1)/(n-p-1) \times (1 - \mathrm{var}(\hat{Y})/\mathrm{var}(Y))$$

Here, $\hat{Y}$ denotes the estimated/fitted abundance.

All the simulations are run under R statistical environment (R Development Core Team, 2008). Each simulation was run for 200 times, and the data sample size was set to 500.

RESULTS
The impact of positive and negative covariances on changing the fitting of explained variation

As shown in Fig. 2 and 3, by controlling the sign of covariance between spatial and environmental variables, we see for pure spatial and environmental variations, a over-fitting problem emerged as the sign is negative (Fig. 2), while an under-fitting problem happened as the covariance sign is positive (Fig. 3). The situations for mixed-effect amount were opposite correspondingly.

For the case of negative covariance simulations (Fig. 2), Welch T-tests suggested that the mean differences between the estimated and true pure variations for both spatial and environmental descriptors are strongly significant ($t=19.27$, $P<2.2e-16$ and $t=18.91$, $P<2.2e-16$ respectively). Further, the same significance level existed for mixed-effect amount too ($t=55.3934$, $P<2.2e-16$). In contrast, the total explained variations between the estimated linear model and designed model are not significantly different ($t=-0.4056$, $P=0.6853$). The results for the case of positive covariances are Similar to those for negative covariances.

The impacts of noise propagation, the strengths of variable coefficients and patterns of spatial/environmental gradients on variation partitioning

As shown in Figs. 4-5, when varying the strengths of coefficients for spatial and environmental variables, there is little impact on the resultant variation explained. Analogously, increasing noise (simulated by increasing the standard deviation) and the generation of spatial and environmental landscapes using different nonlinear equations, the results are similar.

The arrow tracking indicated that, as noise influence level increased (Fig. 6), there is a trend that the estimated explained variation for [SE] will approach the true variations (the red line). However, this situation does not occur in the cases for [S] and [E]. Thus, the results suggested that in real ecological surveying environment, high-level noise magnitude will be very suitable to extract the true co-variation information shared by both spatial and environmental variables. This result also suggested high noise levels in the real environment will not influence the over-fitting or under-fitting problem in variation partitioning.

Conclusively, it seemed that covariance between spatial and environmental factors is the major factor influencing the problem of over-fitting and under-fitting.

DISCUSSION

What's the mixed co-variation?

The most interesting finding derived from our study was that, the co-variation, the overlapping of spatial and environmental variation in true models, is relevant to the covariance (with negative sign) between the spatial and evironmental variables, which reads $-2abCov(S,E)$. Although in estimated models, it should have additional terms (but as we assume covariance is very small, the second-power terms can be omitted), this simple term will let us understand the impact of covariance of spatial and environmental variables on influencing resultant partitioning patterns.

Thus, in true models without estimation when $Cov(S,E) > 0$, the estimated mixed co-variation should be positive, while the real co-variation negative. When $Cov(S,E) < 0$, the estimated mixed co-variation should be negative and the real one positive (Here we always assumed that the coefficients $a$ and $b \geq 0$, see Appendix I Theorem 3 for details). As it is not possible to allow negative variance, but it occurs in the variation partitioning, therefore, we suggested the terms "mixed-effect scalar amount", or "mixed covariation", instead of "mixed variance" in the whole text. Here for consistence, we used "covariation" for the whole text.

This part of covariation has some interesting behaviors. As shown in right-bottom subplots of Figs 2-5, the covariation changed in a way quite different from other parts of variations (e.g., pure environmental/spatial variations). The changing track seemed orthogonal to the red line (*y=x*). This pattern was not changed when we relaxed the setting of parameters $\alpha$ and $\beta$. So, why did it show this line-shifting pattern?

This orthogonal pattern demonstrated an important result about the true co-variation and estimated co-variation. They are negatively proportional in principle, especially when the covariance between variables is much lower than the self-variance of each variable. Their relationship is $[SE]_{real} \approx -[SE]_{est}$ (Theorem 3). When true covariation is increased in the full model from negative to positive, the estimated covariation will be decreased correspondingly from positive to negative, and vice versa. Appendix I provided the analytical solution of this argument.

Therefore, in all previous literature, the co-variation was never correctly calculated and explained. As showed in Theorem 3, the signs for estimated covariation and the observed/real covariation are completely opposite. Therefore, the conventional three-step calculation of mixed-covariation should be adjusted by adding a sign ahead the estimated covariation.

This pattern can be further verified in Fig. 6. Increasing noise levels will make the inference of mixed covariance being highly accurate.

Can we accurately estimate the variations explained by sole spatial or environmental factors?

The answer is yes, but we don't need to follow the three-step procedure completely, as it will cause over-fitting or under-fitting problem. As we have illustrated in the Appendix 1, as long as there is a covariance among independent variables, the fitted coefficients for each variables will depart from the true ones.

To take into account the impact of covariance on over-fitting and under-fitting, I propose an adjusted method for performing variance partitioning, here is the solution,

We have to only consider the full model as follows:

$$Y = aS + bE + \delta$$

As the data are large enough, then,

$$Y \approx aS + bE$$

Thus,

$$Var(Y) = a^2 Var(S) + b^2 Var(E) + 2ab Cov(S, E)$$

The variance explained by spatial factor S and environmental factor E respectivley, thus is,

$$Var(\hat{Y}_S) = a^2 Var(S)$$

$$Var(\hat{Y}_E) = b^2 Var(E)$$

The mixed variance explained by both factors is,

$$MixVar(Y_{S \cap E}) = -2ab Cov(S, E)$$

Thus, it seems not necessary at all to perform three-step methods to partition variations for spatial and environmental variables; instead, one step is enough. The merit of this single regression analysis is that the total explained variation is almost identical to the true total explained variation (e.g., the left-bottom subplot in Figs. 2-4)

CONCLUSIONS

Here by using numerical simulations and mathematical deduction, we addressed the issue that why variation partitioning methods can't accurately predict the true variations contributed by each group of independent variables. We found that three-step variation partitioning methods have the inherent problems to fit the true model, due to the covariance of environmental and spatial variables. This phenomena will occur for any kind of partial regressions. To correct over-fitting and under-fitting problems, I propose that a full regression analysis is enough to obtain all the three-part variations, or it might be not necessary to introduce the mixed-variation as it was directly influenced by the estimation bias of variations for each part of independent variables.

ACKNOWLEDGEMENTS

This work was supported by a UBC graduate scholarship. I very appreciate Prof. Dr. Diane Srivastava and Sally Otto for their comments and instructions.

FIGURES AND TABLES

Fig. 1. A schematic map showing different components and fractions that are used in variation partitioning. [E]-pure environmental variation; [SE]-mixed environmental and space co-variation; [S]-pure space variation; [D]-unexplained variation. [S]+[SE]+[E]=[S+E] denotes the total explained variance.

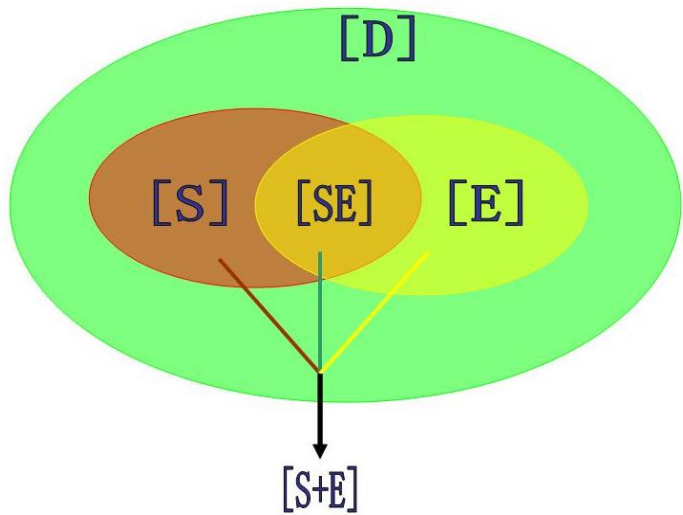

Fig. 2. True and estimated variation comparison, in this case, the covariance among spatial and environmental gradients is set always larger than zero (mean $Cov(S,E) = 0.08$, minimal $Cov(S,E) = 0.015$, and maximal $Cov(S,E) = 0.16$), while the noise was kept in constant ($\sim N(0,1)$). The red line indicates the points where estimated and true variation is consistent. In this case, the over-fitting problem for pure spatial and environmental variations emerges; correspondingly, under-fitting problem was occurred for MEA. All used adjusted-$R^2$ values. Other parameters used for the simulation is $\alpha = 0.27$, $\beta = 10$, $a \sim N(3,1)$ and $b \sim N(7,1)$.

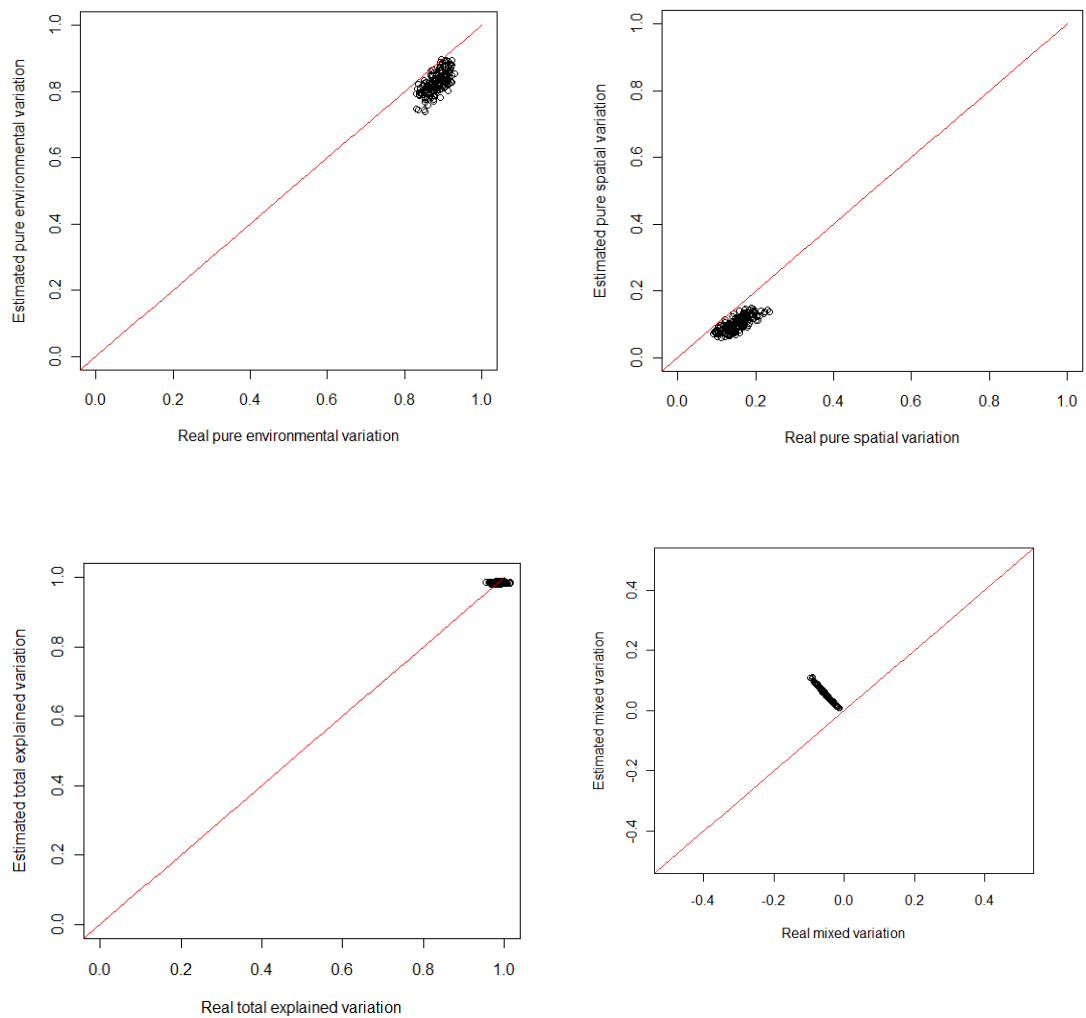

Fig. 3. True and estimated variation comparison, in this case, the covariance among spatial and environmental gradients is mostly negative correlated (mean $Cov(S,E) = -0.08$, minimal $Cov(S,E) = -0.15$, and maximal $Cov(S,E) = 0.008$). Noise mean was kept as zero, while the standard derivation was increased gradually when simulation number increased (minimal S.D.=0, maximal S.D.=2). The red line indicates the points where estimated and true variation is consistent. In this case, the over-fitting problem for pure spatial and environmental variations emerges; correspondingly, under-fitting problem was occurred for MEA. All used adjusted-$R^2$ values. Other parameters used for the simulation is $\alpha = 0.27$, $\beta = 10$, $a \sim U(3,4)$ and $b \sim U(7,8)$.

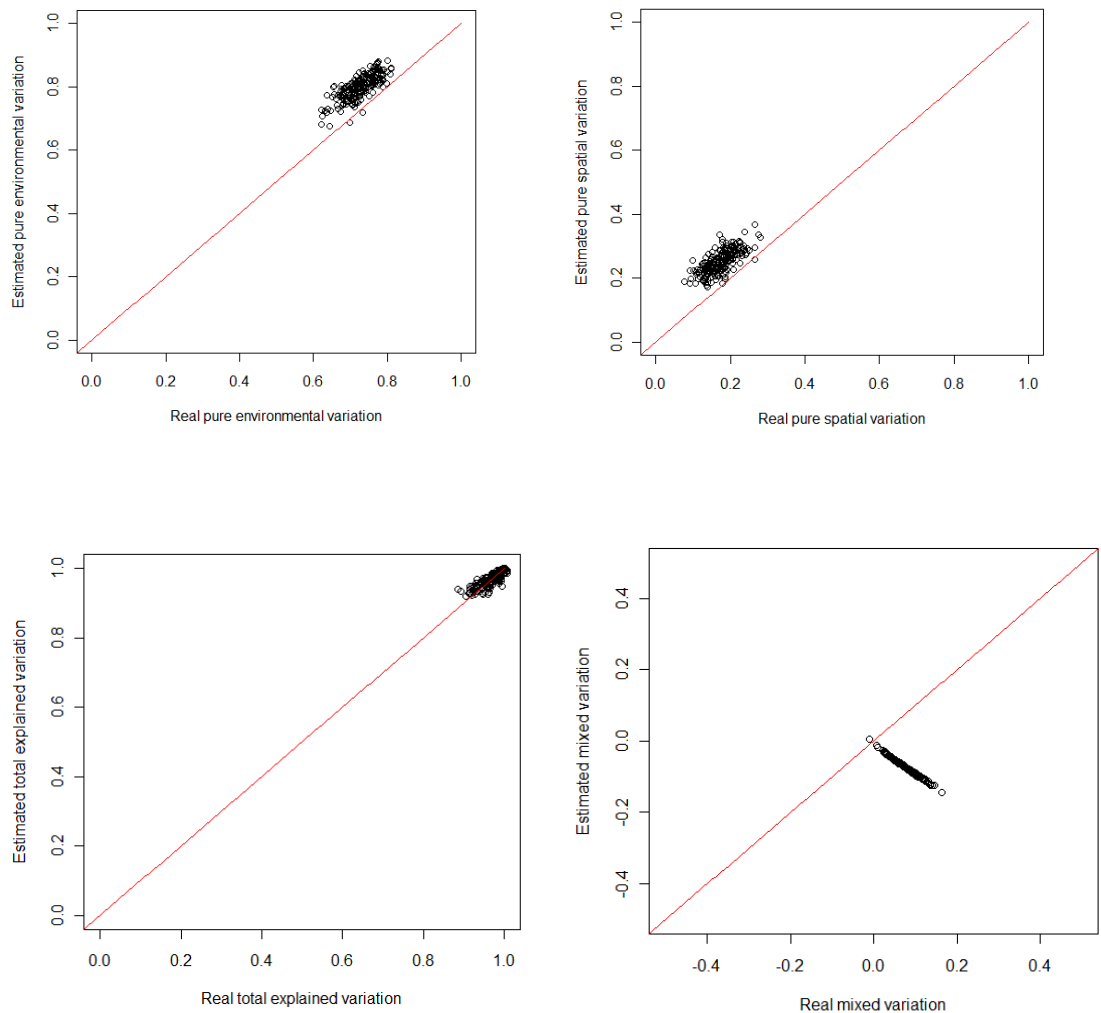

Fig. 4 Increasing the coefficient value ($a$) for spatial variable and its impacts on over-fitting and under-fitting problems in variation partitioning. For each part of variations, increasing the strength of spatial predictor will reduce the quadratic difference between true and estimated variations. All the decreasing trend lines are significant. The squares (red, green, blue respectively) in the last subplot indicated the simulations when $a=3$, $a=4.5$ and $a=6$ respectively. Other parameters: $b=3$, $\alpha=0.27$, $\beta=10$, and $\delta=1$.

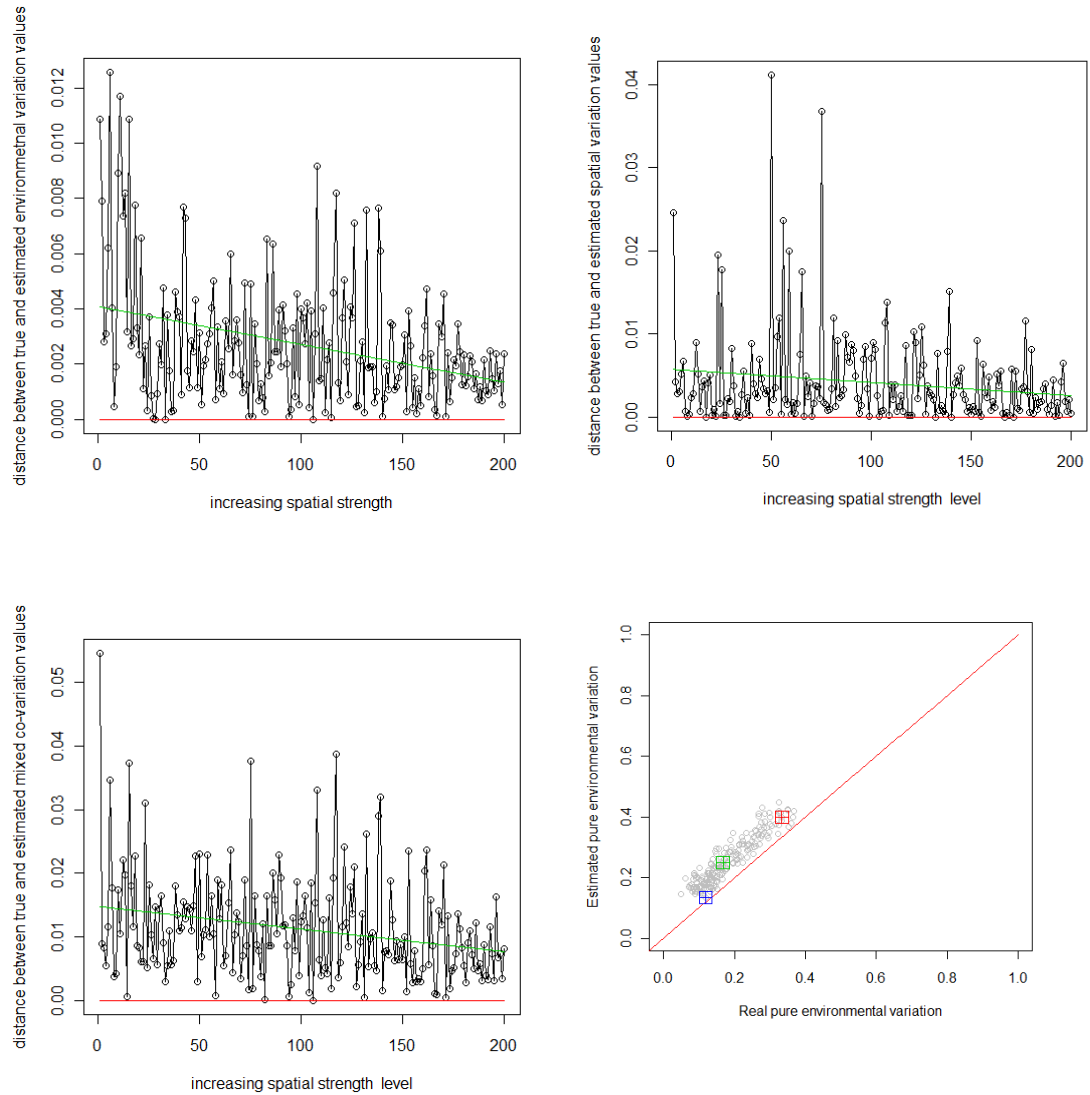

Fig. 5 Increasing the generating coefficient ($\alpha$) for spatial gradient along the landscape locations and its impacts on over-fitting and under-fitting problems in variation partitioning. For each part of variations, increasing the strength of spatial predictor will reduce the quadratic difference

between true and estimated variations. All the decreasing trend lines are significant. The squares (red, green, blue respectively) in the last subplot indicated the simulations when $\alpha = 0.27$, $\alpha = 0.4$ and $\alpha = 0.52$ respectively. Other parameters: $a = b = 3$, $\beta = 10$, and $\delta = 1$.

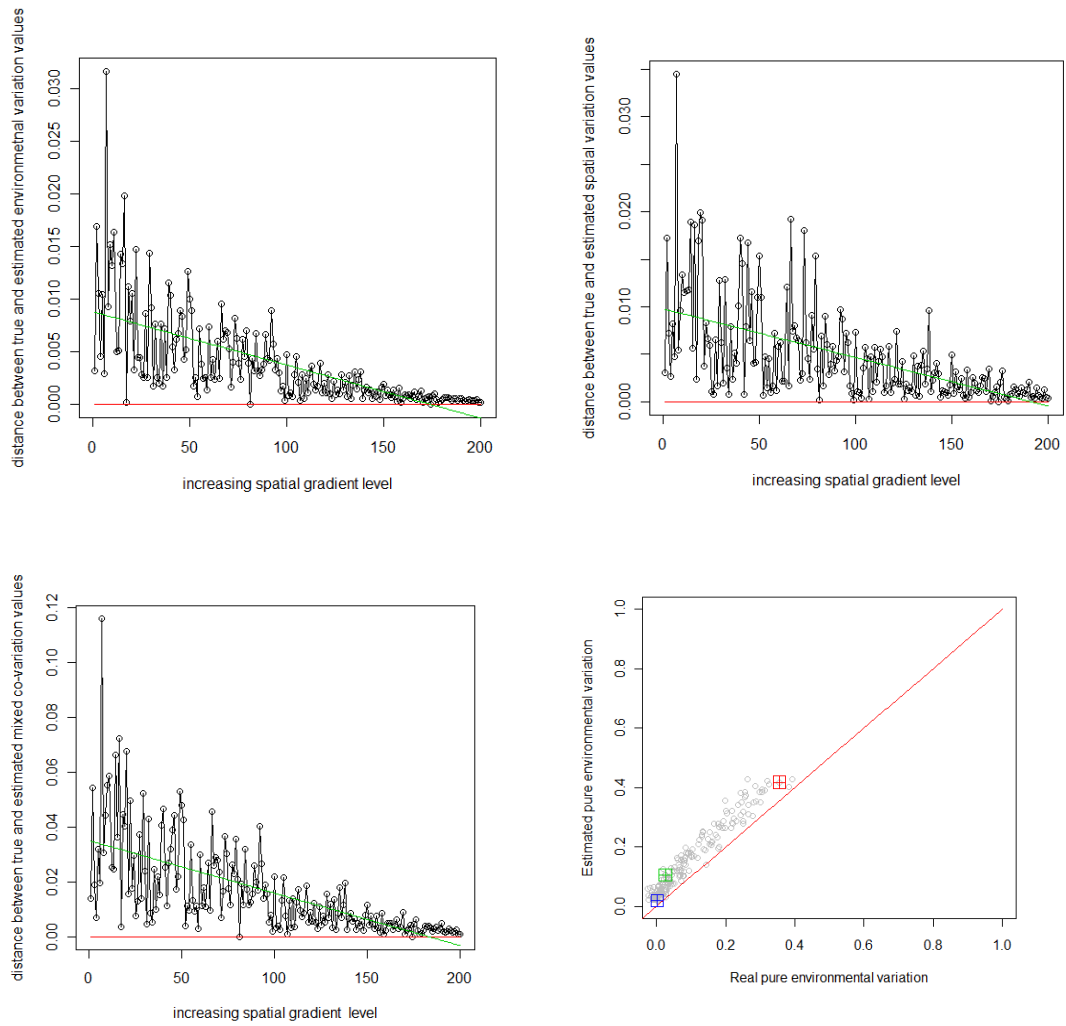

Fig. 6 Increasing the white noise level ($\delta$) and its impacts on over-fitting and under-fitting problems in variation partitioning. For each part of variations, increasing the strength of spatial predictor will reduce the quadratic difference between true and estimated variations. All the decreasing green trend lines are significant (T-test of coefficients). The squares (red, green, blue respectively) in the last subplot indicated the simulations when $SD_\delta = 0.1$, $SD_\delta = 2$ and $SD_\delta = 4$ respectively. Other parameters: $a = b = 3$, $\alpha = 0.27$, and $\beta = 10$.

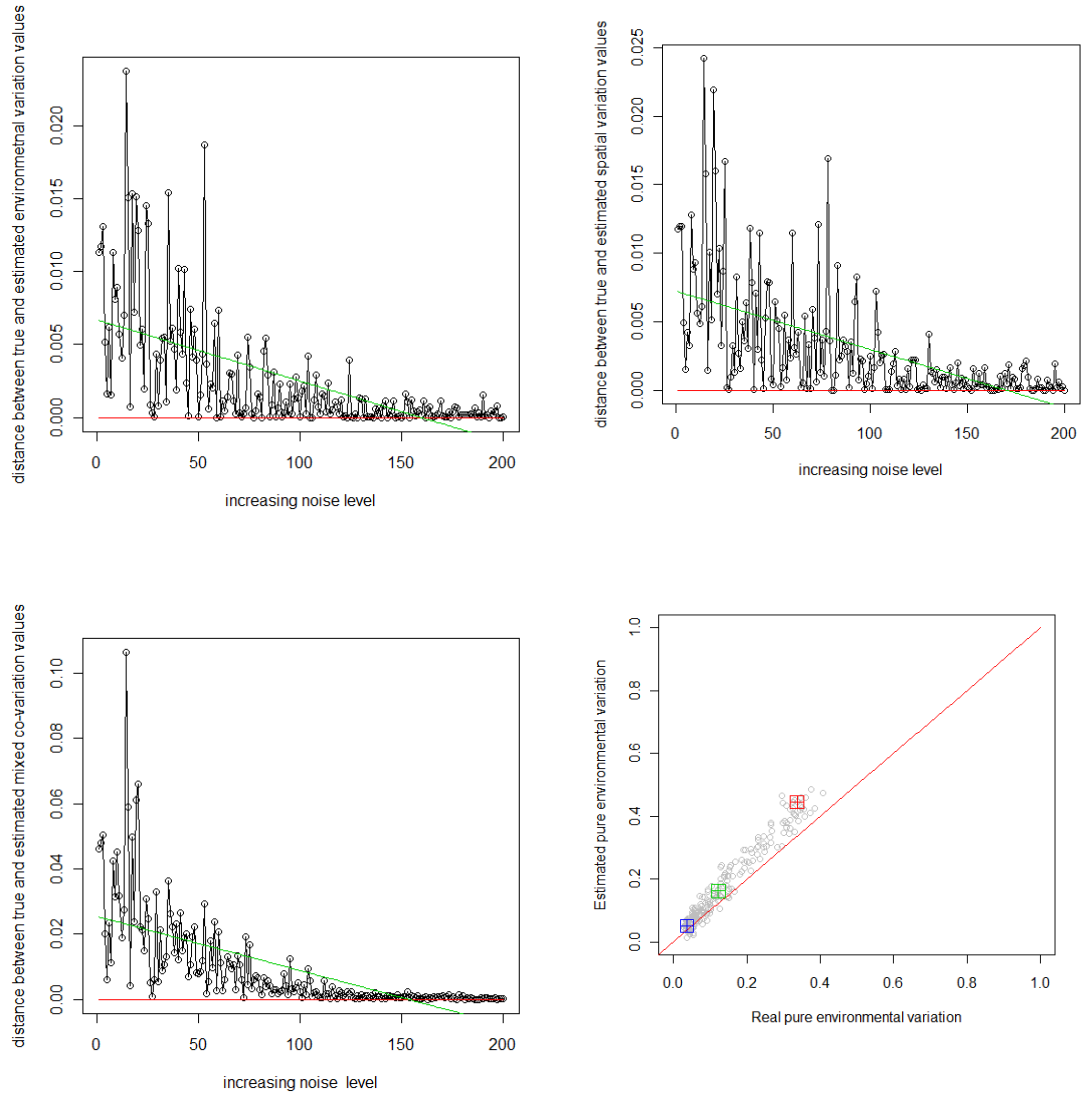

Fig. 7 Full congruence between analytical mixed covariation and numerically simulated mixed covariation.

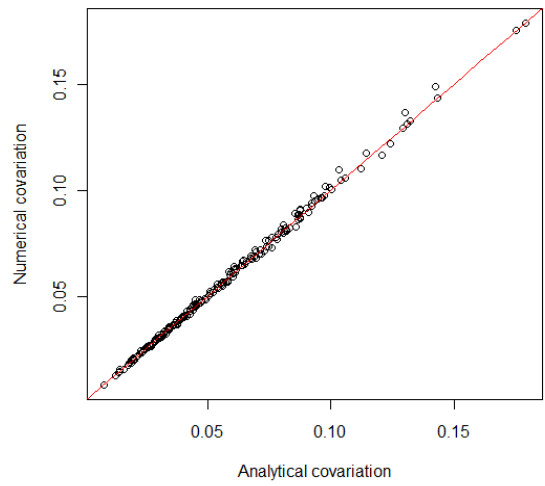

APPENDIX: mathematical deduction of overfitting and underfitting problems in three-step variation partitioning methods

The full model when both spatial and environmental descriptors are necessary predictors should be,

$$Y = aS + bE + \delta$$

Here $\delta \sim N(0,1)$. $a, b$ are the real coefficients, and we always set $a \geq 0$ and $b \geq 0$ in the present paper (if not satisfied, we can simply change the signs of S and/or E) for spatial and environmental variables respectively to generate the response variable Y.

The full model has the total variance as,

$$Var(Y) = a^2 Var(S) + b^2 Var(E) + 2ab Cov(S,E) + 1$$

If we use only environmental or spatial factor as predictors, then we have,

$$\hat{Y} = \hat{a}S$$

$$\hat{Y} = \hat{b}E$$

Thereby,

$$Var(\hat{Y}_S) = \hat{a}^2 Var(S)$$

$$Var(\hat{Y}_E) = \hat{b}^2 Var(E)$$

Here, $\hat{a}, \hat{b}$ are partial coefficients respectively for using sole spatial and environmental predictors respectively.

**Theorem 1:**

If $Cov(S,E) > 0$, we must have, $\begin{cases} \hat{a} > a \\ \hat{b} > b \end{cases}$.

In contrast,

if $Cov(S,E) < 0$, we must have, $\begin{cases} \hat{a} < a \\ \hat{b} < b \end{cases}$.

*Proof:*

If we only want to use spatial descriptor S as the sole predictor, we have,

$$\hat{Y} = \hat{a}S \qquad (2)$$

Our target is to find a suitable coefficient $\hat{a}$ to minimize the quadratic difference between original Y and predicted $\hat{Y}$ (derived from S, E and $\delta$),

$$\underset{\hat{a}}{Arg}\min[(Y-\hat{Y})^2]$$

$$\Rightarrow \underset{\hat{a}}{Arg}\min[Var((a-\hat{a})S+bE+\delta)]$$

$$\Rightarrow \underset{\hat{a}}{Arg}\min[(a-\hat{a})^2 Var(S)+b^2 Var(E)+$$
$$2(a-\hat{a})bCov(S,E)+\delta^2+$$
$$2(a-\hat{a})Cov(S,\delta)+2bCov(E,\delta)]$$

If we write $a-\hat{a}=x$ and $L(x)=Var[xS+bE+\delta]$

Then, for minimizing L(x), we can take the first derivative of L(x) against x, so we get,

$$\frac{dL}{dx}=2xVar(S)+2bCov(S,E)+2Cov(S,\delta)=0$$

So,

$$x=\frac{-bCov(S,E)-Cov(S,\delta)}{Var(S)}$$

Leading to,

$$\hat{a}=a+b\frac{Cov(S,E)}{Var(S)}+\frac{Cov(S,\delta)}{Var(S)}$$

Analogously, for using environmental variable E as the only predictor, we can solve the coefficient as,

$$\hat{b}=b+a\frac{Cov(S,E)}{Var(E)}+\frac{Cov(E,\delta)}{Var(E)}$$

As we assumed that the random variable is independent to observed variables $E$ and $S$ respectively, then we have,

$$\begin{cases} \hat{a}=a+b\dfrac{Cov(S,E)}{Var(S)} \\ \hat{b}=b+a\dfrac{Cov(S,E)}{Var(E)} \end{cases} \quad (3)$$

Thus, when $Cov(S,E)>0$, we have $\begin{cases}\hat{a}>a\\\hat{b}>b\end{cases}$; when $Cov(S,E)<0$, we have $\begin{cases}\hat{a}<a\\\hat{b}<b\end{cases}$.

For the mixed co-variation co-contributed by both spatial and environmental variables, based on the formulation in previous work (e.g., Peres-Neto et al., 2006), it could be written as,

$$CoVar(\hat{Y}_{S\cap E})=(Var(\hat{Y}_S)+Var(\hat{Y}_E))-Var(\hat{Y}_{S\cup E})$$

We used the term $CoVar$ indicated that the mixed covariation can be negative or positive, and different from separated variations for each independent group of variables.

Thus, the above equation characterized the true mixed variance explained by both spatial and environmental variables.

In practice, in full regression analysis, the true total explained variance must be,

$$[S+E] = Var(\hat{Y}_{S \cup E}) = Var(Y_{S \cup E}) = a^2 Var(S) + b^2 Var(E) + 2ab Cov(S,E)$$

In partial regression analysis, since only spatial or environmental descriptor is used for regression, thus, the solution in equation (2) for $\hat{a}$ and $\hat{b}$ is used here, then, we can deduce the equation for mixed variance as follows,

$$[SE]_{est} = CoVar(\hat{Y}_{S \cap E}) = \hat{a}^2 Var(S) + \hat{b}^2 Var(E) - (a^2 Var(S) + b^2 Var(E) + 2ab Cov(S,E))$$
$$= 2ab Cov(S,E) + \frac{a^2 Cov(S,E)^2}{Var(S)} + \frac{b^2 Cov(S,E)^2}{Var(E)} \quad (4)$$

However, in real case, we don't know the influence of another variable, then we should have,

$$[SE]_{real} = CoVar(Y_{S \cap E}) = a^2 Var(S) + b^2 Var(E) - (a^2 Var(S) + b^2 Var(E) + 2ab Cov(S,E))$$
$$= -2ab Cov(S,E) \quad (5)$$

There, we found that, (1) it is very hard to correctly estimate the real mixed-covariation ($CoVar(Y_{S \cap E})$ or $[SE]_{real}$) as well (only in special cases, e.g., $Cov(S,E) = 0$); (2) more importantly, the true and estimated covariations are totally sign-opposite (equations (3) and (4)). This conclusion therefore was arranged as a theorem in Appendix II. The reason for that should be due to the over/under-fitting problems in the tree-step calculation procedure as seemed below.

**Theorem 2:** Positive covariance/correlation between independent spatial and environmental variables will lead to the situation that real covariation was negative, but the estimated covariation was positive. In contrast, negative covariance between the variables will lead to positive real covariation, but the sign of estimated covariation is unknown.

**Proof:**
From equations (3) and (4), it is therefore straightforward to have,

$$[SE]_{real} = CoVar(Y_{S \cap E}) = -2ab Cov(S,E) \quad (5)$$

$$[SE]_{est} = 2ab Cov(S,E) + \frac{a^2 Cov(S,E)^2}{Var(S)} + \frac{b^2 Cov(S,E)^2}{Var(E)} \quad (6)$$

When $Cov(S,E) > 0$, then $[SE]_{real} < 0$ and $[SE]_{est} > 0$, ture covariaton was over-fitted;

when $Cov(S,E) < 0$, then $[SE]_{real} > 0$ and the sign of $[SE]_{est}$ depends on the three terms on the right-hand side. And the fitting status is unknown.

So, let's we look at the difference between the true and estimated covariation between the two equations (3) and (4), we have,

$$CoVar(\hat{Y}_{S \cap E}) - CoVar(Y_{S \cap E})$$
$$= (\hat{a}^2 Var(S) + \hat{b}^2 Var(E)) - (a^2 Var(S) + b^2 Var(E)) \quad (7)$$
$$= (\hat{a}^2 - a^2)Var(S) + (\hat{b}^2 - b^2)Var(E)$$

When $a$ and $b$ is over-estimated (i.e., $Cov(S,E) > 0$; from the theorem in Appendix I) then, the mixed covariation derived from S and E is over-estimated as well ( $CoVar(\hat{Y}_{S \cap E}) - CoVar(Y_{S \cap E}) > 0$ ). The signs Correspondingly, when the pure spatial/environmental variations were under-estimated ( $Cov(S,E) < 0$ ), true $CoVar(Y_{S \cap E})$ was under-fitted accordingly ( $CoVar(\hat{Y}_{S \cap E}) - CoVar(Y_{S \cap E}) < 0$ ). As $CoVar(Y_{S \cap E})$ or $[SE]_{real} > 0$, therefore the sign for $[SE]_{est}$ can be either positive or negative. The fitting status can be explicitly understood, leaving the sign unknown.

However, as in our study, we assumed that the covariance between independent variables was greatly smaller than self variances for each independent group of variables; therefore, the contribution of second-power terms can be omitted, leading to $[SE]_{est} \approx 2abCov(S,E)$. So, in reality for many cases, the sign for estimated covariation $[SE]_{est}$ should be negative (when $Cov(S,E) < 0$).

**Theorem 3**: The signs between true covariation and estimated covariation are totally opposite. And their relationship can be indicated as:

$$[SE]_{real} \approx -[SE]_{est} \quad (8)$$

Sampling bias will not influence this equality basically, the most appealing conclusion is the estimated covariation identified by Borcard and Legendre's method (1992) could not be used as the estimation of ture covariation directly. Our finding showed that a negative sign must be added ahead the estimated covariation value!

APPENDIX II. True and estimated variation comparison under more parameter-relaxed case. The covariance among spatial and environmental gradients is mostly negative correlated (mean $Cov(S,E) = -5.89$, minimal $Cov(S,E) = -68.67$, and maximal $Cov(S,E) = 16.72$).

Noise mean was kept as zero, while the standard derivation was increased gradually when simulation number increased (minimal S.D.=0, maximal S.D.=2). The red line indicates the points where estimated and true variation is consistent. In this case, the over-fitting problem for pure spatial and environmental variations emerges; correspondingly, under-fitting problem was occurred for co-variation. All used adjusted-$R^2$ values. Other parameters used for the simulation is $\alpha \sim U(0.27, 1.27)$, $\beta \sim U(10, 20)$, $a \sim U(3, 4)$ and $b \sim U(7, 8)$.

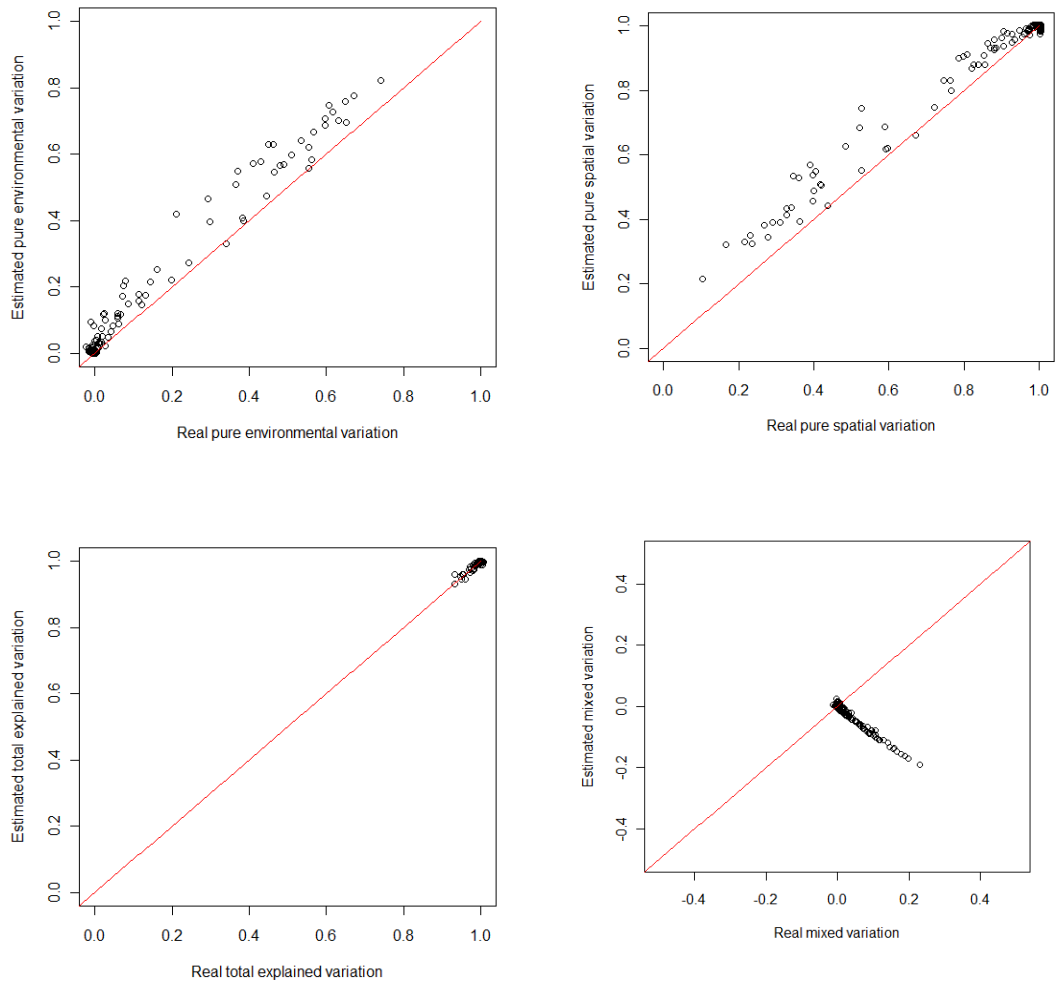